# XULIA- Sistema de control integral de dispositivos con Windows™ diseñado para personas con tetraplejia.

Sistema de accesibilidad

## *XULIA- Comprehensive control system for Windows™ devices designed for people with tetraplegia.*

Accessibility system

Dr. Antonio Losada
anlosada@uvigo.es
Universidad de Vigo

## 1.1 Abstract

XULIA is a comprehensive control system for Windows computers designed specifically to be used by quadriplegic people or people who do not have the ability to move their upper limbs accurately. XULIA allows you to manage all the functions necessary to control all Windows functions using only your voice. As a voice-to-text transcription system, it uses completely free modules combining the Windows SAPI voice recognition libraries for command recognition with Google's cloud-based voice recognition systems indirectly through a Google Chrome browser, which allows you to use Google's paid voice-to-text transcription services completely free of charge. XULIA manages multiple grammars simultaneously with automatic activation to ensure that the set of commands to be recognized is reduced to a minimum at all times, which allows false positives in command recognition to be reduced to a minimum.

## 1.2 Resúmen

XULIA es un sistema de control integral de ordenadores con Windows diseñado específicamente para ser usado por personas tetrapléjicas o personas que no tienen la capacidad de mover con precisión sus extremidades superiores. XULIA permite gestionar todas las funciones necesarias para controlar todas las funciones de Windows empleando únicamente la voz. Como sistema de transcripción de voz a texto emplea módulos totalmente gratuitos conjugando las librerías SAPI de reconocimiento de voz de Windows para el reconocimiento de comandos con los sistemas de reconocimiento de voz de Google en la nube de forma indirecta a través de un navegador Google Chrome, lo que permite emplear de forma totalmente gratuita los servicios de transcripción de voz a texto de Google que son de pago. XULIA gestiona múltiples gramáticas simultáneamente con activación automática para garantizar que en todo momento el conjunto de comandos a reconocer se reduzca al mínimo, lo que permite reducir al mínimo los falsos positivos en el reconocimiento de comandos.

## 1.3 Introducción

En cuanto los sistemas informáticos comenzaron a salir de los laboratorios y a extenderse, fueron siendo necesarios mejores interfaces de interacción con los ordenadores que permitieran alcanzar mayor productividad. En un principio se estandarizó el uso del teclado como único dispositivo de interacción y años después se incorporó el ratón, conjuntamente con la difusión de los interfaces gráficos. Tanto teclado como ratón, han permanecido hasta la actualidad como los dispositivos más usados en los interfaces con los sistemas informáticos. En los últimos años se han desarrollado dispositivos de interacción específica para determinadas necesidades o profesiones y dado el incremento de potencia de los sistemas se ha popularizado tanto el audio como los videos, así que hoy todo dispositivo de cómputo electrónico de uso generalista posee capacidad de gestionar tanto audio como videos, tanto en grabación como reproducción.

Con el acceso de las personas discapacitadas a los ordenadores, ha sido necesario desarrollar interfaces específicos para ellos, dado que en muchos casos tienen inutilizados los miembros que permiten emplear las interfaces estándar. Sobre este punto hay infinidad de alternativas, comenzando con sistemas muy simples con pulsadores como [1] y llegando a los sistemas más complejos para las discapacidades más extremas en las que no es posible mover ningún músculo, en las que pueden emplearse interfaces cerebro-ordenador conocidos como BCI (*Brain Computer*

*Interface*) como en [2] o interfaces con lectura directa de las señales nerviosas que van a los músculos o EMG (*ElectroMyoGraphy*) como en [3]. En otros casos menos extremos en los que pueden mover la cabeza, se emplean sistemas de detección de la posición de la cabeza para mover un ratón como en [4][5][6] e incluso esta tecnología puede emplearse en el sector médico en técnicas en las que las manos están ocupadas y se necesita mover algún elemento con otros miembros como en [7].

En otros casos, en los que mantienen la capacidad de mover la lengua, pero no son capaces de emitir sonidos claros, también se han desarrollados interfaces que detectan el movimiento de la lengua mediante la incorporación de un imán a la misma, que se lee mediante unos sensores colocados a ambos lados de la cara como en [8].

Las lesiones y afectaciones pueden ser infinitas, y en algunos casos es posible que mantengan cierta capacidad de movilidad de sus manos, pero no con la destreza suficiente como para poder emplearlas con normalidad. En estos casos también se han desarrollado interfaces hombre-máquina que permiten detectar ciertas posiciones o gestos de las manos para interactuar con los sistemas como en [9] [10] o en [11] donde el sistema permite manejar una silla de ruedas.

Pero los únicos sistemas que se han implantado de modo masivo son los que emplean lenguaje natural, dado que pueden ser utilizados con ciertas modificaciones por personas con discapacidad como se puede ver en [12] [13], pero también son una forma muy natural de comunicarse con un ordenador para personas que no presentan ningún tipo de discapacidad.

La evolución de estos sistemas se lleva realizando desde los años 50 como se ve en [14], pero los primeros trabajos aplicados se comenzaron a ver en los años 80 con sistemas como [15] del año 1986 que permitía la grabación e identificación de comandos de voz y estaba especialmente diseñado para personas con discapacidad o [16] del mismo año, que permitía realizar encuestas telefónicas. En [17] [18] del año 1991 se extendían las capacidades pudiendo grabar y reconocer frases. En el año 2001 en trabajos como [19] se comienza a ver el control de dispositivos físicos mediante la voz, dado que las tasas de error en reconocimiento de comandos comienzan a ser aceptables. En [20] se comienzan a ver las primeras aplicaciones de reconocimiento de voz en dispositivos móviles como las antiguas PDAs (*Personal Digital Assistant*).

A partir del año 2002 la capacidad de los dispositivos personales ya era suficiente para ejecutar software complejo que requería extensos recursos computaciones y se comenzaron a desarrollar nuevos modelos como describen [21] y [22]. Grandes empresas como IBM y Microsoft comenzaron a desarrollar sus propios sistemas de reconocimiento de voz, a los que posteriormente se les unieron Google, Amazon, Philips, etc.

En la Sección 6.4 se citan los motores de reconocimiento de voz que se encontraban disponibles en el año 2014, año en el que se desarrolló Xulia, y en la Sección 6.5 se detallan las aplicaciones de control de sistemas Windows® que existían en ese mismo año.

## 1.4 Descripción general de Xulia como sistema de accesibilidad

Hace muchos años que el ser humano ha comenzado a realizar progresos para comunicarse mediante lenguaje natural con los ordenadores. Esta funcionalidad que inicialmente representa una mejora en los interfaces de comunicación, que resultan en una interface más ágil para transmitir comandos a nuestros ordenadores, para algunas personas con discapacidad, son el único o uno de los pocos interfaces que pueden ser empleados para comunicarse con el ordenador.

El sistema que se presenta en este trabajo ha sido desarrollado específicamente para ser empleado por personas tetrapléjicas. Como objetivo, se pretende alcanzar, únicamente empleando la voz, el nivel de productividad de una persona sin limitaciones físicas que emplee teclado y ratón, en el manejo de un ordenador con sistema operativo Windows. Adicionalmente, se pretende implantar este sistema en Brasil, con lo que será necesario disponer de tecnología de conversión de voz a texto en lenguaje portugués, como mínimo en el modo de dictado continuo. En la especificación de requisitos debe tenerse en cuenta que la operativa del sistema no debe necesitar en ninguna circunstancia el empleo del teclado, ratón o cualquier otro dispositivo que deba ser empleado con las manos.

Cada año, en España se producen 22 nuevos casos de lesiones medulares por cada millón de habitantes, lo que supone 1.000 personas por año. Según la OMS, a nivel mundial se contabilizan medio millón de nuevos lesionados medulares por año. De entre todas las lesiones, la categoría más frecuente es la tetraplejia incompleta (el 40.6%) seguido por paraplejia incompleta (el 18.7%), paraplejia completa (el 18.0%), y tetraplejia completa (el 11.6%). En el tipo de lesión más común, la tetraplejia incompleta, los afectados, normalmente no pueden mover o emplear las manos para realizar movimientos que requieran precisión.

En este tipo de lesiones se produce la parálisis total o parcial tanto de los miembros superiores como de los inferiores. Normalmente se ocasiona por traumas producidos en la médula espinal en su zona superior, pero en otros muchos casos, viene ocasionado por enfermedades que provocan degeneración de las neuronas motoras, como en el caso de la ELA (Esclerosis Lateral Amiotrófica). Debemos tener en cuenta que la vida de una persona con tetraplejia cambia radicalmente. Este signo clínico, junto con los presentados por enfermedades neuronales, originan que estos pacientes puedan alcanzar grados de dependencia cercanos al 100%, conservando sus funciones mentales totalmente intactas.

Toda lesión medular que origina tetraplejia debe producirse en la zona superior de la médula espinal, entre las vértebras C4 y C7. Cuanto más abajo se produce la lesión, mejor pronóstico tendrá y mayor grado de autonomía. Sólo en el caso de lesiones producidas en la C4 podrá tener problemas de dicción, con lo que el sistema de reconocimiento de voz podría tener problemas con la calidad del reconocimiento. Con el resto de las lesiones, los afectados podrán hablar de modo normal.

Debemos tener en cuenta que el sistema debe ser totalmente autónomo y debe estar diseñado para sustituir de modo completo las funciones del teclado y el ratón en todo momento, dado que la voz debe ser la única interface que puede emplear el usuario.

XULIA nace en el año 2015 como una iniciativa dentro del *Instituto Novo Ser* de Brasil [189]. Este instituto desarrolla actividades para mejorar la calidad de vida de las personas discapacitadas con movilidad reducida. Ha sido una de las entidades impulsoras para la creación de Motrix [188] en la Universidad de Rio de Janeiro en el año 2002. Motrix es un sistema gratuito, de control integral de un ordenador con Windows® empleando la voz, pero en la época en la que fue desarrollado, Microsoft no tenía la tecnología de reconocimiento continuo de voz, por lo que solo es capaz de reconocer comandos independientes. Adicionalmente, al estar basado en las librerías de reconocimiento de voz de Microsoft en su versión SAPI 4, solo reconoce idioma inglés. Motrix cumple sus objetivos, pero una de sus carencias más graves está relacionada con el dictado continuo. Cuando el usuario quería dictar un texto, tenía que hacerlo letra a letra, y dado que el sistema presentaba un porcentaje alto de falsos positivos con comandos muy cortos, cada letra se correspondía con una palabra. Por ejemplo, para dictar hola, el usuario tenía que decir: hotel, oscar, lima, alfa. Como se puede ver, el trabajo de dictar textos grandes era totalmente agotador, además de extremadamente lento.

Desde el Instituto Novo Ser estaban solicitando la creación de un nuevo software que permitiera dictado continuo de voz en idioma portugués empleando recursos gratuitos, ya que pretendían poder llegar a todas las clases sociales. En el año de creación de Xulia, la única tecnología de reconocimiento de voz que soportaba el dictado continuo en idioma portugués era la de Google, pero los sistemas gratuitos que Google ponía a disposición de los usuarios solo estaban disponibles en el sistema operativo Android®. Dentro de Windows® había la posibilidad de emplear el servicio en la nube de reconocimiento de voz de Google, pero éste tenía un coste de 0,012 céntimos de euro por cada 10 segundos de audio procesado. Partiendo de este dato se calcula un coste mensual de 100 € para un uso continuo de 3 horas al día, lo que supone un importe a tener en cuenta en países como Brasil, en donde su moneda está muy devaluada y un gran porcentaje de su población tiene pocos recursos.

Inicialmente tampoco nos planteamos emplear un sistema operativo Android como servidor de reconocimiento de voz, ni en modo físico ni en virtual, ya que la configuración del sistema y el número de elementos sería muy alta y en caso de querer virtualizar una máquina Android, la potencia del sistema de emulación debía ser muy alta.

Aparte de la tecnología de reconocimiento de voz de Google tenemos otras dos grandes empresas con aplicaciones de escritorio, como Microsoft y Nuance (con su *Dragon Dictate* y cuya tecnología de reconocimiento de voz es la base de la de *Siri*). Entre ellas, solo Microsoft ofrece un servicio básico de reconocimiento de voz gratuito dentro de Windows. En el año 2015 Microsoft tenía publicada la versión 5 de sus librerías de reconocimiento de voz fuera de línea, pero éstas sólo soportaban los idiomas inglés, francés, alemán, español, chino y japonés, con lo que el portugués no estaba soportado.

Adicionalmente en las pruebas del reconocimiento continuo de voz con sus librerías SAPI 5, con gramáticas abiertas de dictado empleando los motores de reconocimiento de voz de Microsoft ofrecidos en Windows, hemos detectado que la calidad del reconocimiento era muy baja, generando un gran número de falsos positivos, sobre todo en presencia de ruido de fondo, con una tasa de casi el 20% de fallos de reconocimiento. Todo ruido de fondo era reconocido como una palabra con un grado de fiabilidad calculado por el sistema bastante alto, cuando realmente representaba un falso positivo, lo que generaba un alto número de errores de reconocimiento. En el caso de emplear micrófonos con una alta atenuación de ruido, los filtros aplicados originaban que el audio resultante perdía características y el sistema reducía el número de falsos positivos, pero también perdía la capacidad de reconocer comandos que no fueran dictados de un modo muy perfecto.

En contrapartida, si no se empleaban las gramáticas de modo dictado, en donde están disponibles para reconocer todas las palabras del idioma, sino que se especificaba la lista de comandos válidos en un momento dado y esta lista era relativamente corta, el sistema de reconocimiento de Windows mejoraba muchísimo su grado de reconocimiento, incluso con ruido de fondo. Como resultado de este estudio inicial llegamos a la conclusión de que, en cada momento, sólo debemos tener activos los comandos mínimos necesarios para que el reconocedor realice la búsqueda sobre el menor número de patrones de voz distintos.

Alternativamente, al probar el sistema de reconocimiento de voz de Google, que solo operaba en dictado continuo, vimos que su sensibilidad al ruido de fondo era muy baja y que el sistema de reconocimiento se comportaba realmente bien en cualquier situación, pero tenía como contrapartida que solo estaría disponible de modo gratuito dentro del sistema operativo Android.

En investigaciones posteriores pudimos comprobar que recientemente (año 2015) Google había incorporado una API Beta de reconocimiento de voz en su navegador Google Chrome. Desde el propio JavaScript y ejecutándose en el propio navegador Chrome era capaz de procesar el audio recuperado del micrófono, lo enviaba a la nube de Google y esta devolvía la traducción de voz a texto del audio recuperado que podía ser tratado desde el navegador con código JavaScript.

Tras realizar varias pruebas comprobando que el funcionamiento era muy correcto, también detectamos varias limitaciones a su uso en modo local. En el momento de activar el reconocimiento de voz desde JavaScript, el reconocimiento de voz solo estaba activo durante unos segundos, tras este tiempo, se desactivaba automáticamente. Cada vez que JavaScript activaba el reconocimiento de voz, aparecía un cuadro de diálogo pidiendo al usuario que permitiera el uso del micrófono. Otra limitación en el uso de esta API, es que no se encontraba disponible en páginas cargadas de modo local desde el sistema de ficheros, lo que nos obligaría siempre a trabajar con servidores web.

Finalmente, tenemos el inconveniente de que el texto reconocido se devolvía al propio JavaScript en el navegador y por razones de seguridad desde JavaScript no pueden emplearse recursos locales.

En la situación actual, teníamos que resolver varios problemas para poder emplear esta tecnología de reconocimiento de voz desde el escritorio de Windows. Inicialmente nuestro objetivo era desarrollar un sistema de control integral de un sistema Windows que interactuara con el sistema simulando las entradas del teclado y ratón, con capacidad de dictado continuo en idioma portugués. Debemos tener en cuenta que nuestro objetivo es permitir que una persona tetrapléjica sea capaz no solo de usar, sino trabajar de modo completo con el ordenador, únicamente con nuestro sistema, sin recursos adicionales ni necesidad de soporte por terceras personas, esto implica que, el sistema no puede quedar bloqueado en ningún momento solicitando la acción del usuario. Para el uso del ordenador, nos hubiera bastado con un sistema que fuera capaz de interpretar un número limitado de comandos, si bien pretendemos que el usuario pueda trabajar de modo completo con el ordenador y esto incluye dictado de correos o elaboración de documentos, para lo cual resultará imprescindible poseer un sistema de reconocimiento continuo de voz en su propio idioma.

En cuanto al requisito del reconocimiento en idioma portugués, viene definido debido a que nuestro usuario de prueba inicial era una persona tetrapléjica que formaba parte del Instituto Novo Ser de Rio de Janeiro en Brasil. El sistema de dictado continuo debería cumplir los siguientes requisitos:

- Por un lado, necesitamos que el sistema esté habilitado de forma permanente, dado que la versión actual del API JavaScript de Chrome se deshabilita al cabo de 10 segundos al no detectar entrada de voz.

- Necesitamos que no pregunte en cada arranque del sistema de reconocimiento si permitimos el uso del micrófono.
- Necesitamos que sea posible cargarlo desde un servidor web instalando el menor número de componentes en nuestro equipo.
- Es necesario que el resultado del reconocimiento sea enviado a una aplicación local para que pueda interactuar con el equipo y enviar esta entrada procesada mediante el API de Windows de bajo nivel a cualquier aplicación abierta simulando que la entrada hubiera sido introducida mediante teclado.
- Necesitamos controlar en remoto el estado de activación del módulo JavaScript que se ejecuta en Chrome para que no se encuentre de forma permanente intentando reconocer audio, sino que solo esté activo cuando se encuentre activa la funcionalidad de reconocimiento continuo de voz.

Para poder implementar estos requisitos, hemos desarrollado un servidor web local que sirve una página dinámicamente modificada con el código JavaScript de reconocimiento de voz, específico para las opciones de configuración actuales. Este servidor publica su servicio en el puerto 8080. En el arranque del sistema de reconocimiento, se pide la página con el código de reconocimiento de voz, junto con la información del idioma a procesar y el estado de arranque automático. Si no se requiere arranque automático, el sistema levanta un navegador Google Chrome con el reconocimiento de voz en modo desactivado. Dado que no queremos retrasos en el momento de activar el dictado continuo, el navegador Chrome está siempre ejecutándose. La primera vez que se ejecuta el API de reconocimiento de voz de Google Chrome, el propio navegador solicita los permisos de uso del micrófono. Una vez otorgados, no los volverá a pedir nunca más, persistiendo esta configuración para siempre. Dado que estamos sirviendo la página desde una dirección local, no será necesario emplear el protocolo SSL para que estos permisos sean persistentes. En caso de servir la página desde un servidor remoto, sería necesario disponer de certificado de servidor y conexión SSL para poder dejar almacenados los permisos de uso del micrófono.

Adicionalmente el propio código del navegador activa un *Timer* que se llama cada segundo con el que se pregunta al servidor, empleando tecnología AJAX, por el siguiente comando o modo de operación al que debe cambiarse. De esta forma, el navegador puede activar o desactivar de modo inmediato la funcionalidad de reconocimiento continuo de voz.

Una vez enviado el comando de inicio de dictado continuo al navegador Chrome, el navegador recupera constantemente el audio, lo envía a la nube de Google y recupera el texto reconocido de retorno. La transferencia del texto reconocido por los servicios de Google se realiza mediante una llamada AJAX desde nuestro navegador Chrome al servidor web local. Este servidor forma parte del sistema de control de Windows que posteriormente procesará esta entrada y nos permitirá manejar el ordenador.

Mediante este módulo de interacción con el API JavaScript de reconocimiento de voz implementado en la API JavaScript de Google Chrome, se consigue emplear los servicios de Google para reconocimiento continuo de voz en cualquiera de los idiomas soportador por Google. En el año 2015 el portugués era uno de los lenguajes soportados, aunque no implementaba la funcionalidad de reconocimiento de los signos de puntuación.

Una vez solucionados los problemas tecnológicos para soportar dictado continuo, debemos desarrollar un sistema base que sea capaz de reconocer comandos del modo más preciso posible y que sea capaz de interactuar a nivel de APIs de sistema con el resto de las aplicaciones y con el propio Windows para posibilitar una experiencia de uso lo más satisfactoria posible. Sumados a estos requisitos, debemos intentar que el grado de productividad de este nuevo sistema que debe operarse solo por voz, sea lo más similar posible al empleo de un ordenador con un teclado y un ratón.

Después de comprobar los puntos débiles del sistema de reconocimiento de voz de Microsoft en su versión SAPI 5, o al menos de la versión gratuita que nos ofrece para el desarrollo de aplicaciones de escritorio de Windows decidimos que debíamos restringir el número de comandos disponibles al mínimo posible en cada momento, con lo que aumentaríamos las posibilidad de que el comando reconocido fuera el correcto y reduciríamos los falsos positivos que aparecen de un modo demasiado habitual al emplear esta tecnología con ruido de fondo en la habitación.

Aunque se ha implementado un modo que permite que los comandos del sistema sean obtenidos empleando la tecnología reconocimiento de voz de Google a través de Chrome, el problema fundamental de este sistema, es que el reconocimiento se efectúa en la nube, lo que implica un tiempo de retardo entre la locución del comando y la obtención del texto del mismo, esto reducía drásticamente el grado de productividad, sobre todo en lo relacionado con comandos que debían de ser inmediatos como los relativos al movimiento de ratón.

Por este motivo, en el modo comando se emplea la tecnología de reconociendo de voz de Windows, lo que limitaba el lenguaje de los comandos al idioma inglés, francés, alemán, español, japonés y chino, en el año 2014.

Desde el año 2017 ya está disponible el idioma portugués brasileño, aunque Microsoft no lo incorporó a la versión de la librería SAPI 5 y para poder emplear este idioma es necesario desarrollar aplicaciones universales Windows o bien migrar las aplicaciones a la versión de servidor de la librería SAPI 10. Más adelante se explicarán los problemas que tienen tanto la versión de servidor de la librería SAPI 10 como la versión de la librería de reconocimiento de voz para aplicaciones UWP (*Universal Windows Platform*).

En cualquiera de las versiones de estas librerías, menos en la versión para UWP (que emplea *Cortana*), aún resulta obligatorio reducir el alto número de falsos positivos que presenta este sistema para lo que se debe mantener el mínimo número de comandos distintos activos en cada momento.

Adicionalmente, el objetivo era crear un sistema multi-idioma, cuyos comandos pudieran ser totalmente configurables y adaptables a cualquier idioma, dado que muchas de las dolencias de nuestros usuarios originan que no sean capaces de emitir de forma correcta ciertos fonemas, que deberán ser eliminados en los comandos cambiándolos por otros.

## 1.5 Técnicas de gestión de la transcripción de audio a texto

La capacidad de reconocimiento de voz la ofrecen las distintas librerías o servicios empleados para el reconocimiento, pero disponer de la traducción a texto del audio de la voz es el primer paso para poder gestionar todas las funciones necesarias para controlar un ordenador y realizar las tareas diarias de forma productiva. Para mitigar los defectos del sistema de reconocimiento y mejorar la experiencia y la productividad

del usuario, se han implementado las siguientes técnicas de gestión de los comandos o del texto reconocido:

- Activación contextual de gramáticas.
- Activación de gramáticas automáticamente por aplicación.
- Desactivación de contextos automáticamente por tiempo.
- Pilas de navegación de gramáticas.
- Comandos de alta seguridad temporizados.
- Multiplicadores de acciones.
- Generación de listas de sustitución en modo dictado.
- Especificación de porcentajes de reconocimiento individual para cada comando.
- Múltiples portapapeles.
- Sistema de posicionamiento del ratón de alta velocidad por rejilla
- Movimientos diagonales del ratón

Sumados al control del modo comando, también tenemos que solucionar la carencia de la implementación de signos de puntuación en el sistema de reconocimiento continuo de algunos idiomas por el sistema de reconocimiento de Google. Para solucionar este último problema se han implementado listas de sustitución automática en modo dictado, de forma que una vez que nuestro sistema detectaba ciertos patrones en modo dictado, automáticamente eran sustituidos por signos de puntuación o caracteres especiales, como el retorno de línea o el salto de página.

Antes de proceder, para comprender los conceptos empleados, resulta básico comprender el concepto de gramática. En Xulia, una gramática es un conjunto de palabras o frases que se encuentran activos para reconocimiento en un momento dado.

A continuación, se describen las siguientes soluciones empleadas para mejorar el comportamiento del sistema.

### 1.5.1 Activación contextual de gramáticas

No todos los comandos se encuentran activos al mismo tiempo, dependiendo del contexto, las gramáticas se activan o desactivan automáticamente o empleando los propios comandos. Por ejemplo, si estamos en modo comando y necesitamos deletrear un texto técnico, dado que no contiene ninguna palabra del lenguaje que estamos empleando dictamos el comando *deletrear*. Esta palabra está asociada a una macro que cambia la gramática activa a la gramática que permite deletrear, con lo que se activan los comandos de dicha gramática, desactivando la gramática activa actualmente. Estos cambios de gramática pueden ser aditivos o sustitutivos. Serán aditivos cuando incrementamos los comandos de la nueva gramática a la gramática actual, y sustitutivos cuando desactivamos los comandos de la gramática actual y activamos solo los de la nueva gramática.

### 1.5.2 Pilas de navegación de gramáticas

Cuando activamos una nueva gramática, esta se coloca en la parte superior de la pila de gramáticas. A partir de ese momento se activarán los comandos de la nueva gramática. Si queremos volver a la gramática anterior, solo tenemos que dictar el comando *volver* y si queremos volver directamente a la gramática inicial de comandos solo tenemos que dictar el comando *modo comando*, este último vacía la pila de gramáticas y activa la gramática de comandos por defecto. Con estas sencillas instrucciones de navegación controlamos el intercambio de gramáticas.

### 1.5.3 Gramáticas de aplicación

Dentro de Xulia, cada aplicación puede tener sus propios comandos, e incluso, un mismo comando puede tener asociadas macros distintas dependiendo de la aplicación activa en cada momento. Estás gramáticas se activan automáticamente, cuando Xulia detecta que una aplicación pasa a primer plano. A partir de ese momento, todos los comandos definidos para ser usados en la aplicación están disponibles y a su vez se desactiva cualquier otra gramática de aplicación que se encontrara activa. Las gramáticas se desactivan automáticamente cuando la aplicación se cierra o cambiamos el contexto a otra aplicación. De este modo conseguimos mantener gramáticas con comandos polimórficos, de modo que un mismo comando realiza acciones distintas dependientes de la aplicación que se encuentra activa en el momento de dictar el comando. Con estos modos de operación conseguimos mantener el número mínimos de comandos activos en cada momento, sin renunciar a tener cientos de comandos específicos para cada aplicación.

### 1.5.4 Comandos protegidos

Xulia permite definir comandos protegidos. Estos comandos solo pueden ser identificados si entramos en modo protegido y una vez dentro de este modo tenemos 3 segundos para dictar el comando antes de que el sistema salga del modo protegido de forma automática. Este modo se emplea para comandos peligrosos o para los comandos de activación y desactivación de Xulia, dado que si desactivamos Xulia accidentalmente perdemos el control de nuestro equipo y dado que este sistema está diseñado para personas con tetraplejia, nuestros usuarios perderían la posibilidad de interactuar con el ordenador y necesitarían de ayuda de otra persona para poder restablecerlo.

### 1.5.5 Cadenas de sustitución

Esta funcionalidad se emplea en modo dictado continuo a través de tecnología Google. Este modo de dictado no está igual de evolucionado en todos los idiomas, esto implica que en algunos lenguajes funciona la identificación de las palabras, pero no funciona la identificación de los signos de puntuación. Con esta funcionalidad podemos elegir cualquier frase para que cuando la encuentre en el texto la sustituya por cualquier otra o por un signo de puntuación. Para poder dictar la frase elegida sin que sea sustituida debemos antecederla de la palabra "literal". De esta forma añadimos la funcionalidad para incorporar tildes, acentos, retornos de línea y demás signos y acciones de control, aunque no las soporte el dictado de Google.

### 1.5.6 Control del ratón relativo y absoluto mediante rejillas

Otra de las características fundamentales para el correcto manejo de nuestro escritorio es el control del ratón. Xulia posee comandos para desplazar el ratón en las cuatro direcciones comunes, pero también permite desplazamiento en las cuatro diagonales. Adicionalmente permite controlar la velocidad de desplazamiento seleccionando entre 4 velocidades, pero donde presenta una solución novedosa es en el desplazamiento absoluto a una posición. Para desplazarse a una posición concreta debemos emplear el comando *rejilla*, con ello XULIA superpone una rejilla con fondo transparente encima de la pantalla con 24 filas y 24 columnas. En cada fila y columna tenemos una letra del alfabeto, para hacer referencia a una celda solo tenemos que especificar el comando asociado a la letra correspondiente a su fila y después a su columna, con lo que el ratón se situará en el medio de la celda indicada.

Con este sistema, sólo con dos palabras podemos posicionarnos de forma inmediata en 576 posiciones, pero con esta precisión no seriamos capaces de clicar directamente en muchos lugares pequeños. Para mejorar esta precisión sin necesidad de dictar múltiples comandos y, sobre todo, sin necesidad de colocar más líneas en la rejilla virtual que entorpecen la visión del fondo, cada una de estas 576 celdas se descompone virtualmente en 9 posiciones, con lo que realmente tenemos 5.184 posiciones para situar de forma directa nuestro ratón. La descomposición es virtual porque tenemos que imaginarlo, ya que si no lo hiciéramos y pintáramos todas las rayas sería complicado ver lo que tenemos debajo de la rejilla. Para activar el modo de posicionamiento de alta precisión, solo tenemos que dictar el número de la celda virtual (1 a 9) antes de dar las dos letras que forman las coordenadas de la celda. Si no dictamos el número antes, el ratón se sitúa en el centro de la celda. Si empleamos los números, el ratón se sitúa en el centro de la subcelda virtual del número correspondiente. Si empleamos el número 0 el ratón se sitúa en la esquina superior izquierda de la celda, con lo que realmente disponemos de 5.760 posiciones distintas de posicionamiento con una sola indicación directa.

### 1.5.7 Multiplicador de comandos

Con Xulia, tenemos una funcionalidad que permite la multiplicación de comandos. Antes de un comando podremos activar el multiplicador dictando la palabra *por* seguida de un número, con ello el siguiente comando se repetirá automáticamente tantas veces como hayamos configurado en el multiplicador. Esta funcionalidad resulta muy interesante para borrados de caracteres o desplazamientos con los cursores.

### 1.5.8 Sistema de configuración y parametrización

Como se ha descrito en puntos anteriores, Xulia es un sistema multi-idioma. Esto quiere decir que los comandos a dictar y las acciones asociados a dichos comandos son totalmente configurables.

Xulia posee un total de 140 macros parametrizadas que podemos elegir asociar a cualquier frase también configurable en cualquier de los idiomas soportador por

Windows. Adicionalmente, el modo dictado continuo de Google puede ser configurado para reconocer cualquiera de los idiomas soportados por los sistemas de reconocimiento de voz de Google y si se establece la configuración para emplear el dictado de Google como fuente de dicción de comandos, también podemos emplear todos sus idiomas en modo comando, aunque perdemos velocidad de reconocimiento, ya que como se ha dicho con anterioridad, el reconocimiento de Google se realiza en la nube.

Para configurar todas estas combinaciones, Xulia tiene una interface gráfica que finalmente genera un fichero de configuración en XML, que los usuarios avanzados podrían modificar directamente. Con este sistema de configuración permitimos reaprovechar partes de los ficheros de configuración entre usuarios o simplemente copiar una configuración de un usuario en otro muy fácilmente.

### 1.5.9 Módulo conversacional

Como funcionalidad adicional, Xulia posee implementado en su interior un chatbot basado en lenguaje AIML (*Artificial Intelligence Mark-up Languaje*). Este lenguaje permite especificar patrones mediante una estructura XML que posteriormente serán buscados en el texto de una conversación para identificar contextos y respuestas. Este lenguaje ha sido especialmente diseñado para crear chatbots. El lenguaje de programación AIML fue desarrollado por el Dr. Richard Wallace y la comunidad de código abierto *Alicebot* entre los años 1995 y 2000. Con él se crearon las bases del primer *Alicebot*, que ha ganado el concurso *Loebner Prize Contest* en el año 2000, 2001, 2004 y también el concurso *The most human* tres veces, y el Campeonato de Chatbot en 2004.

Para interpretar los tags AIML se ha empleado el intérprete en C# Program# [190] desarrollado por Nicholas H.Tollervey, que hemos modificado para poder incorporar comandos parametrizables con estructura compleja dentro de las los patrones reconocidos por el intérprete AIML.

De entrada, el cerebro AIML de Xulia contiene un total de 34.000 patrones y está basado en el cerebro de Alexia que ha sido creada en la Universidad de Vigo dentro del proyecto Galaia.

Su cerebro gestiona múltiples temáticas de conversación como son: relaciones personales, política, hobbies, ciencia, cine, música, animales, apariencia física, astrología, psicología, deporte, dinero, drogas, edad, estados de ánimo, familia, filosofía, humor, informática, profesiones, religión, etc.

### 1.5.10 Modos de operación

Xulia tiene los siguientes modos de operación:

- Modo comando.
- Modo conversación.
- Modo dictado continuo Google.
- Modo dictado continuo Windows.
- Modo rejilla.
- Modo corrección.
- Modo deletrear.

El comportamiento de cada uno de los modos de operación es el siguiente:

**Modo comando**

En este modo, Xulia emplea la tecnología de reconocimiento de voz integrada dentro de Windows para aplicaciones de escritorio. La ventaja de este modo es la velocidad de reconocimiento, necesaria para comandos de interacción con el escritorio y comandos de posicionamiento y desplazamiento del ratón. A la par tiene dos grandes desventajas, por un lado, presente una tasa alta de falsos positivos en presencia de ruido de fondo, cuando el número de palaras a reconocer es muy numerosa. Adicionalmente otro problema que presenta es que inicialmente, esta tecnología solo funciona en los lenguajes inglés, francés, alemán, español, chino y japonés. Posteriormente fue introducido el portugués, pero esta limitación nos obliga a escoger el idioma más similar en otros países.

En este modo podemos configurar cualquier palabra o frase de los idiomas citados y asociarlos a una secuencia parametrizada y combinada de las 140 macros predefinidas.

**Modo conversación**

En este modo se activa el chatbot Xulia y pasaremos a estar conversando con la inteligencia artificial generada en AIML. Entre las funcionalidades incluidas se encuentra la posibilidad de mantener una conversación con seguimiento de contexto, pero adicionalmente, dado que se ha extendido el lenguaje AIML para dar cabida a patrones de ejecución de comandos, podremos ejecutar comandos complejos parametrizados con frases dictadas en lenguaje natural. Algunas de las funciones que pueden realizarse con estos comandos complejos son las siguientes:

- Buscar cualquier contenido en internet.
  - En esta funcionalidad, Xulia abrirá directamente un navegador con el resultado de la búsqueda en el buscador predeterminado.
- Preguntar directamente por *el tiempo* de cualquier ciudad.
  - Xulia buscará directamente la temperatura, grado de humedad, velocidad del viento y precipitaciones, interpretando el resultado y leyendo el pronóstico mediante voz sintetizada.
- Preguntar por *quién e*s una persona.
  - En caso de encontrar el resultado, lo interpretará y leerá la respuesta.
- Preguntar *por qué es*.
  - Si encuentra el resultado, lo interpretará y leerá la respuesta.
- Enviar un correo a una dirección de la lista de contactos.
  - Permitirá especificar el asunto y el contenido de un mensaje de correo, que enviará directamente a su destinatario.
- Gestionar la lista de la compra.
  - Permite tanto incluir nuevos productos, como crear y eliminar listas de compra.

**Modo dictado continuo Google**

Mediante este modo se activa el sistema de reconocimiento indirecto a través de un navegador Google Chrome, de forma que Xulia emplea los sistemas de reconocimiento de voz en la nube de Google.

En este modo empleamos dos características importantes de Xulia:

- Cadenas de sustitución.
- Ventana de corrección.

Con las cadenas de sustitución podemos asociar a un comando expresiones recurrentes, de forma que serán escritas directamente dictando el comando asociado, aunque la utilidad más importante es emplearlas para caracteres de puntuación y caracteres de control. El dictado Google no está igual de avanzado en todos los idiomas, y en nuestro caso, en idioma portugués en el momento del desarrollo de Xulia, todavía no tenía soportados los caracteres de control ni los signos de puntuación, con lo que al dictar la palabra *punto*, el sistema reconocía la palabra *punto* o al dictar *nueva línea* igualmente lo reconocía de modo literal. Con las listas de sustitución podemos asociar los caracteres que queramos a frases, de modo que si las reconoce dentro del texto las sustituirá por los caracteres asociados. En caso de que queramos evitar esta transformación, debemos anteceder la frase de la palabra *literal*.

A pesar de que la calidad del reconocimiento es muy buena, siempre se mantendrá un porcentaje de fallos que deben ser corregidos, y en nuestro caso debemos tener en cuenta que este sistema está diseñado para personas tetrapléjicas, con lo que el sistema de corrección debe funcionar de forma muy rápida empleando comandos de voz.

**Modo corrección**

El usuario puede cambiar en cualquier momento al modo corrección. Al activar este modo, Xulia muestra la ventana de corrección y muestra la última fase dictada. El sistema separa las palabras de la frase y las numera (ver Figura 7.3.1). De este modo, el usuario puede seleccionar el número de la palabra con la que quiere operar para después corregirla o eliminarla. Una vez aceptada una corrección se sustituye el resultado en la aplicación en la que se estaba dictando.

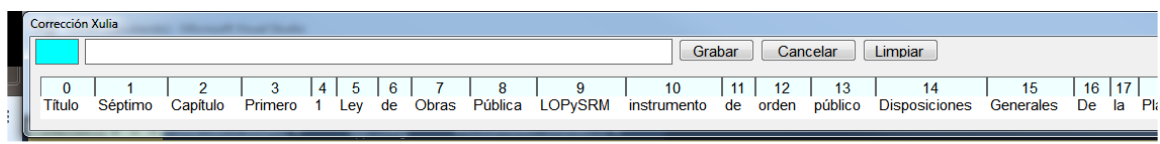

*Figura 7.3.1. Imagen de la ventana de corrección con todas las palabras de la frase numeradas.*

## 1.6 Versiones de las librerías de reconocimiento de voz empleadas en XULIA

Inicialmente, desde su creación, XULIA ha empleado la versión 5 de las librerías SAPI (*Speech Application Program Interface*). Estas librerías permitían tanto el reconocimiento de frases y palabras seleccionadas como la gramática de dictado libre, aunque con una calidad bastante baja en entornos ruidosos, al menos en lo relativo al dictado continuo. Adicionalmente se emplea la tecnología de reconocimiento de voz de Google en la nube de un modo indirecto mediante el empleo de un navegador Google Chrome como servidor de reconocimiento de voz.

En el año 2017 se ha realizado una actualización en las librerías empleadas para realizar el reconcomiendo de voz. Actualmente, y desde el año 2017 tenemos disponibles tres tecnologías de reconcomiendo de voz de Microsoft. Cada una de ellas emplean distintas APIs y no son compatibles entre sí, aunque la versión 5 y la versión 10 tienen grandes similitudes, aunque ambas no soportan las mismas funciones.

Desde la creación de XULIA, se ha empleado la versión 5 de las librerías SAPI de reconocimiento de voz. Este motor de reconocimiento posee la capacidad de reconocer palabras concretas indicadas en una gramática de reconocimiento, pero también posee el modo de dictado continuo, en donde reconoce conversación normal en cualquiera de los 6 idiomas que soporta. Este modo está disponible desde el año 2000. Estas librerías se publican bajo el espacio de nombres *System.Speech.*

Con posterioridad Microsoft publico la versión 10 de sus librerías de reconocimiento de voz. El motor que implementa esta API es *Microsoft Server Speech.* Este API permite únicamente el reconocimiento de palabras pertenecientes a una gramática prefijada, dado que el motor de reconocimiento no implementa el dictado continuo. La calidad de reconocimiento es algo superior al motor de SAPI 5, pero en presencia de ruido de fondo no representa un salto trascendente. La ventaja de este sistema es que ha ampliado el número de idiomas soportados a 26, incluyendo desde el año 2018 el idioma portugués. Estas librerías se publican bajo el espacio de nombres *Microsoft.Speech.*

Por último, desde la aparición de Cortana en Windows 10, Microsoft ha diseñado un nuevo motor de reconocimiento de voz muy superior a los dos anteriores. La tecnología de este nuevo motor es la misma que emplea Cortana. Este motor ha sido diseñado para ser empleado en las aplicaciones UWP (*Universal Windows Platform*) de Windows. Este motor soporta reconocimiento de palabras y frases prefijadas fuera de línea con un alto grado de inmunidad al ruido, y dictado continúo empleando los recursos de la nube de Microsoft, aunque de modo gratuito. Xulia, al interactuar a medio nivel con las librerías de Windows para controlar el sistema, no puede reconvertirse a una aplicación UWP, sin embargo, mediante enlaces dinámicos de librerías, es posible emplear este motor de reconocimiento en aplicaciones de escritorio. Sin embargo, este motor que es, con mucha diferencia el más avanzado actualmente de Microsoft, para nuestro propósito, posee un gran inconveniente que lo hace inoperativo. El motor necesita que la aplicación que lo esté empleando permanezca siempre con el foco en primer plano. En cuando la aplicación pierde el foco (deja de ser la aplicación activa), el motor desactiva el sistema de reconocimiento. Debemos tener en cuenta que XULIA es una aplicación que trabaja en segundo plano y cuyas acciones se reflejan en otras aplicaciones con las que está trabajando el usuario, lo que nos impide emplear esta tecnología. Estas librerías se publican bajo el espacio de nombres *Windows.Media.Speech.* Con la última modificación realizada sobre Xulia, se han combinado las funcionalidades de las tres librerías. Se emplea SAPI 10 para reconocimiento de comandos y frases prefijadas con posibilidad de emplear 26 idiomas y cuando se cambia a modo dictado local empleamos SAPI 5 soportando dictado en 6 idiomas.

Finalmente se ha añadido un nuevo modo de dictado con la tecnología de las librerías UWP que abre una nueva ventana y permite dictar de modo continúo empleando el nuevo motor de Windows 10. Todo el texto reconocido se muestra en la pantalla y una vez que se cierra, el contenido de la ventana se copia al portapapeles para pegarlo en cualquier aplicación.

## 1.7 Motores de reconocimiento de voz de alto rendimiento multilenguaje

En el año 2015, antes del inicio del desarrollo de XULIA, se analizaron todas las alternativas que permitieran reconocimiento continuo de voz, intentando identificar aquellos sistemas gratuitos con capacidad de reconocimiento de idioma portugués. En los siguientes apartados se describen los sistemas analizados.

### 1.7.1 Microsoft

Mediante las librerías SAPI (*Speech Application Programer Interface*), Microsoft incorporó la posibilidad de añadir la funcionalidad de reconocimiento de voz en aplicaciones Windows de forma gratuita.

Las primeras versiones de SAPI 1 se lanzaron en el año 1995 con Windows 95. Estas versiones poseían la capacidad de reconocer palabras o frases independientes prefijadas. No fue hasta la versión 3, donde se incorporó el dictado discreto y en la versión 4 se permitía, con ciertos problemas, reconocer dictado continuo en inglés, que fue empleado por varias aplicaciones como [23]. Estas versiones no son compatibles con .NET

En el año 2000 se lanza SAPI 5, en la que Microsoft rediseñó por completo el sistema de reconocimiento de voz, añadiendo la capacidad de reconocer inglés, francés, alemán, español, chino y japonés, con capacidad de dictado continuo con cierta calidad. Esta es una de las versiones del API que se emplea en Xulia para el dictado continuo fuera de línea, aunque presenta demasiada sensibilidad al ruido de fondo y a la calidad de los micrófonos utilizados. Para emplear estas librerías desde .NET debe emplearse el espacio de nombres *Microsoft.Speech*.

Posteriormente, Microsoft lanza la versión 10.0 de su sistema de reconocimiento de voz para servidor bajo el espacio de nombres *Microsoft.Speech*. Esta versión no era totalmente compatible con la anterior y tenía como principal inconveniente que no implementaba el dictado continuo, pero como contrapartida soportaba el reconocimiento de palabras o frases prefijadas en un total de 26 idiomas, entre los que se encontraba el idioma portugués.

Con el lanzamiento de Windows 10, Microsoft dio un salto de gigante en su sistema de reconocimiento de voz, mejorando uno de los grandes problemas que siempre mantuvo, que era la alta sensibilidad al ruido de fondo, con el que generaba falsos positivos. Finalmente, con Windows 10 y su asistente Cortana, el reconocimiento de voz era casi inmune al ruido de fondo y la calidad del sistema pasó a otro nivel. Al mismo tiempo, Microsoft lanzó su sistema de desarrollo de aplicaciones universales (UWP) con el que se podía crear una única aplicación y ejecutar en múltiples plataformas. El problema fue que incorporó muchas limitaciones para el desarrollo

UWP. Para impulsar esta plataforma, las nuevas APIs de reconocimiento de voz con la potencia mejorada solo fueron publicadas para aplicaciones UWP y adicionalmente tenían una grave limitación. El nuevo sistema de reconocimiento solo permanece activo mientras la aplicación que lanza el reconocimiento está activa en primer plano. Solo por esta limitación, se hace imposible poder emplear estas nuevas APIs en una aplicación como Xulia, que debe recuperar la información del usuario para actuar sobre otras aplicaciones que se encuentran en primer plano.

Adicionalmente, con la mejora de la calidad del reconocimiento se debe pagar un precio. Con el API UWP el reconocimiento de palabras y frases prefijadas se realiza fuera de línea, pero el reconocimiento del dictado continuo se lleva a cabo en los servidores de Microsoft en la nube, por lo que es necesario disponer en todo momento de conexión a internet.

### 1.7.2 Microsoft Cloud

Este es un servicio de pago de Microsoft incluido en la plataforma Microsoft Azure. En la Tabla 7.5.1 se pueden ver sus costes. Este Servicio convierte voz a texto con mucha mayor precisión comparado con las librerías de Windows. Este software de última generación admite más de 85 idiomas globales junto con variantes. Puede personalizar modelos agregando palabras específicas y mejorar la precisión de su texto para frases específicas de dominio.

Esta solución admite entradas de audio de varias fuentes, como archivos de audio, almacenamiento de blobs y micrófonos. Puede utilizar el registro del hablante para determinar las palabras exactas, y también obtiene transcripciones altamente legibles automáticamente con puntuación y formato.

*Tabla 7.5.1. Coste del empleo de los modelos en la nube de Microsoft.*

| | |
|---|---|
| Estándar | €0,844 por hora de audio |
| Custom Speech | €1,181 por hora de audio |
| Hospedaje de puntos de conexión de Custom Speech | €33,74 por modelo y mes |
| Audio multicanal de transcripción de conversaciones | N/D por hora de audio[4] |

### 1.7.3 Nuance

Antes del desarrollo de *ViaVoice*, IBM desarrolló un producto llamado *VoiceType*. En 1997, *ViaVoice* se presentó por primera vez al público en general. Dos años después, en 1999, IBM lanzó una versión gratuita de *ViaVoice* [24]. En 2003, IBM otorgó a ScanSoft, que poseía el producto competitivo *Dragon Naturally Speaking*, los derechos exclusivos de distribución global de los productos *ViaVoice* Desktop para Windows y Mac OS X. Dos años más tarde, Nuance se fusionó con ScanSoft. La fusión de todas estas tecnologías dio lugar años más tarde a *Dragon Naturally Speaking* que es uno de las mejores aplicaciones de escritorio de pago de reconocimiento de voz. Pero las fusiones e incorporaciones tecnológicas no cesaron, dado que en año 2011, Nuance adquirió otra gran compañía con tecnología de reconocimiento y síntesis de voz como es Loquendo.

La última compra finalizó en marzo de 2022, donde Microsoft adquiere Nuance. Las librerías de Nuance son la base de uno de los chatbots más evolucionados hoy en día llamado **SIRI** presente en la plataforma iOS de Apple.

### 1.7.4 Google Cloud

Los sistemas de reconocimiento de voz de Google, actualmente son de los mejores. En el año 2017, Google conseguía un 95% en su tasa de reconocimiento para el inglés de los Estados Unidos; el más alto de todos los asistentes de voz actualmente disponibles. Esto se traduce en una tasa de error de palabras del 4,9%, lo que hace que Google sea el primero del grupo en caer por debajo del umbral del 5%.

En la tabla 7.5.2 pueden verse los costes de utilización de los servicios de reconocimiento de voz en la nube de Google.

*Tabla 7.5.2. Coste de los servicios en la nube de Google.*

| Función | Modelos estándar (todos excepto los modelos mejorados para vídeo y teléfono) | | Modelos premium* (modelos mejorados para vídeo y teléfono) | |
|---|---|---|---|---|
| Tiempo | De 0 a 60 minutos | De más de 60 minutos a 1 millón de minutos | De 0 a 60 minutos | De más de 60 minutos a 1 millón de minutos |
| Reconocimiento de voz (sin el almacenamiento de registros de datos) | Gratis | 0,006 USD por cada 15 segundos** | Gratis | 0,009 USD por cada 15 segundos** |
| Reconocimiento de voz (con el almacenamiento de registros de datos) | Gratis | 0,004 USD por cada 15 segundos** | Gratis | 0,006 USD por cada 15 segundos** |

### 1.7.5 Amazon

Amazon tiene un servicio en el que unifica el reconocimiento de voz junto con las capacidades conversacionales de su chatbot Alexa, por lo que actualmente no tiene un servicio de reconocimiento de voz puro independiente.

### 1.7.6 IBM

El servicio de reconocimiento de voz en la nube de IBM es *Watson Speech to Text*. Es una solución avanzada de reconocimiento y transcripción de voz que funciona con inteligencia artificial y permite una transcripción rápida y precisa en varios idiomas y casos de uso, incluidos el análisis de voz y la interpretación del lenguaje natural para agentes de asistencia. Trabaja mediante sofisticados modelos de aprendizaje automático. Mediante un coste adicional, permite realizar entrenamiento adicional de los modelos, con voz proporcionada por el cliente. En la tabla 7.5.3 se detallan sus costes.

*Tabla 7.5.3. Precios de los servicios en la nube de reconocimiento de voz de IBM.*

| Minutos de reconocimiento* | Coste/minuto |
|---|---|
| 1 – 250000 | 0,0200 $ |
| 250001 – 500000 | 0,0150 $ |
| 500001 – 1000000 | 0,0125 $ |
| 1000000 + | 0,0100 $ |

*$0,03 adicional por minuto al emplear modelos *Custom Language Model Add-on*.

### 1.7.7 Mozilla DeepSpeech

Estas librerías son de código abierto, pero solo soportan por el momento el idioma inglés, y se encuentran en una etapa temprana de desarrollo.

DeepSpeech [25] es un motor de voz a texto de código abierto que utiliza un modelo entrenado con técnicas de aprendizaje automático. Emplea en su implementación las librerías TensorFlow de Google. Esta plataforma de código abierto está diseñada para la decodificación avanzada con una integración flexible del conocimiento. El núcleo del sistema es una red neuronal recurrente bidireccional (BRNN) entrenada para ingerir espectrogramas de voz y generar transcripciones de texto. Por el momento está disponible para su uso un modelo preentrenado en idioma inglés.

El software necesita Python 2.7 y Git Large File Storage (una extensión de Git para crear versiones de archivos grandes).

### 1.7.8 Julius

Julius [26] es otro motor de reconocimiento con licencia Open Source. Actualmente tiene disponibles de forma gratuita los modelos de lenguaje para los idiomas japonés e inglés.

Julius es un motor de reconocimiento continuo de voz (LVCSR) de dos pasos y alto rendimiento. Está basado en trigramas y trabaja con modelos ocultos de Márkov realizando decodificaciones en tiempo real. Los modelos acústicos y los modelos de lenguaje son conectables, y puede crear varios tipos de sistemas de reconocimiento de voz creando sus propios modelos y módulos para que se adapten a distintas tareas. El motor principal se implementa como una biblioteca integrable, con el objetivo de ofrecer la capacidad de reconocimiento de voz a diversas aplicaciones.

### 1.7.9 Kaldi

Kaldi [27] es un conjunto de herramientas de reconocimiento de voz de última generación escrito en C++. Está destinado a ser utilizado principalmente para la investigación de modelado acústico.

Kaldi proporciona un sistema de reconocimiento de voz basado en transductores de estado finito (usando la librería OpenFst disponible gratuitamente), junto con documentación detallada y scripts para construir sistemas de reconocimiento completos. Su biblioteca central admite el modelado de tamaños arbitrarios de contexto fonético, modelado acústico con modelos de mezcla gaussiana subespaciales (SGMM), así como modelos de mezcla gaussiana estándar, junto con todas las transformaciones lineales y afines de uso común.

Aunque el software ha estado en desarrollo durante algunos años, se encuentra en una etapa temprana de desarrollo y no hay lanzamientos oficiales del software. Actualmente en su web pueden descargarse versiones iniciales de modelos en inglés y chino mandarín.

### 1.7.10 Philips SpeechLive

Philips tiene su propio servicio de reconocimiento de voz en la nube llamado *SpeechLive* [28] con varias opciones, como el procesado de ficheros, reconocimiento mediante grabación por voz y servicios en tiempo real de traducción. En su página no da detalles técnicos del servicio, pero indica que tiene una precisión cercana al 99%. Por el momento solo tiene modelos de los idiomas inglés, alemán y francés cuyos costes pueden verse en la Tabla 7.5.4.

*Tabla 7.5.4. Costes de los modelos de Philips.*

| Modelo de idioma | Coste/minuto |
|---|---|
| Inglés | 1,25 € |
| Alemán | 2,54 € |
| Francés | 3,09 € |

## 1.8 Aplicaciones de control de Windows con la voz

En el año 2014, cuando se inició el desarrollo de Xulia, existían varias aplicaciones de control de Windows mediante la voz. Debe tenerse en cuenta que Xulia se desarrolló por petición del Instituto Novo Ser de Brasil, con el requisito de soportar el reconocimiento continuo de voz en portugués. Ninguna de estas aplicaciones soportaba este idioma para reconocimiento continuo.

### 1.8.1 Windows Speech Recognition

WSR *(Windows Speech Recognition)* es un sistema de control integral de Windows mediante el reconocimiento voz. Fue desarrollado por primera vez para Windows Vista. Permite manejar todas las funciones de Windows empleando únicamente la voz, sustituyendo de modo completo el teclado y el ratón. También permite dictar textos, navegar por la web y posicionarse de modo directo en cualquier lugar de la pantalla empleando una rejilla dinámica.

El motor de reconocimiento de voz se ejecuta en local, no siendo necesario que el ordenador se encuentre conectado a internet. Adicionalmente permite incorporar nuevas palabras dictadas por el usuario al diccionario.

Permite reconocimiento de comandos independientes y también permite dictado continuo. La precisión del reconocimiento es muy buena con comandos definidos, pero al emplear la gramática para dictado continuo presenta muchos falsos positivos originados por el ruido de fondo, por lo que resulta imprescindible emplearlo con micrófonos de muy buena calidad con atenuación de ruido de fondo. WSR viene incorporado con Windows sin coste adicional.

En sus últimas versiones que datan del año 2018, soporta los idiomas inglés, francés, alemán, español, portugués brasileño, japonés y chino.

### 1.8.2 Dragon Naturally Speaking

*Dragon Naturally Speaking*[1] es un software de reconocimiento de voz desarrollado inicialmente por Dragon Systems basado en modelos ocultos de Markov. Actualmente pertenece a *Nuance Communications* y su versión actual es la 15. Sus últimas versiones emplean redes neuronales recurrentes entrenadas mediante aprendizaje profundo.

Posee versiones tanto para Windows como para iOS. Permite controlar completamente todas las funciones de la interface, posibilitando el dictado continuo sobre la aplicación activa.

---

[1] Fuente: https://www.nuance.com/es-es/dragon.html [Acceso: 8-may-2022]

También permite definir nuevos comandos y crear macros para realizar acciones complejas. La edición personal en su versión 15 tiene un coste de 499 $ y soporta los idiomas inglés, francés, alemán, español, holandés y japonés.

### 1.8.3 Braina

Braina[2] (Brain Artificial) es un asistente personal inteligente. Es un software de inteligencia artificial que permite interactuar con el ordenador mediante comandos de voz en multitud de idiomas. Es un software de pago. Una vez analizados sus requisitos, asumimos que emplea la misma tecnología que Xulia para utilizar los sistemas de reconocimiento de voz de Google en la nube empleando un navegador Google Chrome como intermediario. Los comandos por defecto están en idioma inglés, siendo necesario cambiar la configuración predeterminada para definir comandos en otros idiomas que solo es posible con las versiones de pago que tienen un coste de 200 $. Sumado a la capacidad de reconocimiento de comandos y de dictado continuo, posee internamente un chatbot, de modo que es posible dictar ciertas frases preconfiguradas en lenguaje natural que serán reconocidas y procesadas por Braina.

### 1.8.4 VozComando

Voxcommando[3] emplea la tecnología de reconocimiento de voz de Microsoft. No está diseñado para prescindir completamente del teclado y el ratón, pero permite definir macros y comandos con los que se puede agilizar y acelerar ciertas tareas en el ordenador.

### 1.8.5 Voice Finger

Voice Finger[4] es un software de reconocimiento de voz que emplea el sistema de reconocimiento de voz de Microsoft. Está diseñado para poder controlar por voz directamente el ratón y el teclado con comandos cortos. Posee un conjunto de comandos muy reducido para sustituir estos periféricos, pero no posee comandos generales directos de control de Windows.

---

[2] Fuente: https://www.brainasoft.com/braina/ [Acceso: 8-may-2022]
[3] Fuente: https://voxcommando.com/home/ [Acceso: 8-may-2022]
[4] Fuente: http://voicefinger.cozendey.com/ [Acceso: 8-may-2022]

## 1.9 Asistentes personales

Los asistentes personales no están diseñados para controlar todas las funciones de un ordenador, pero sí pueden acelerar tareas, permitiendo que se den las instrucciones en lenguaje natural. En los siguientes apartados se describen los asistentes personales más conocidos.

### 1.9.1 Siri

Siri es un asistente virtual que forma parte de los sistemas operativos de Apple. Utiliza consultas de voz, soporta el seguimiento de contexto y posee una interfaz de usuario de lenguaje natural para responder preguntas, hacer recomendaciones y realizar acciones. Sus servicios de reconocimiento de voz y procesamiento del lenguaje natural se encuentran en la nube. Con el uso continuo, se adapta a los usos, búsquedas y preferencias de idioma individuales de los usuarios, brindando resultados individualizados.

La tecnología de su motor de reconocimiento de voz fue desarrollada, en sus orígenes, por *Nuance Communications*. Actualmente emplea tecnologías avanzadas de aprendizaje automático.

Siri admite una amplia gama de comandos de usuario, incluida la realización de llamadas telefónicas, la consulta de información básica, la programación de eventos y recordatorios, el manejo de la configuración del dispositivo, la búsqueda en Internet, la búsqueda de información sobre entretenimiento y puede interactuar con aplicaciones integradas en iOS.

### 1.9.2 Cortana

Cortana es un asistente virtual desarrollado por Microsoft que emplea la plataforma Azure de servicios en la nube de Microsoft para el reconocimiento de voz y el procesamiento del lenguaje natural. En los comandos de búsqueda emplea el motor Bing. Actualmente, Cortana está disponible en inglés, portugués, francés, alemán, italiano, español, chino y japonés.

Lleva el nombre de Cortana, por el personaje de inteligencia sintética de la franquicia de videojuegos Halo de Microsoft.

### 1.9.3 Google Assistant

Google Assistant es un asistente virtual impulsado por inteligencia artificial desarrollado por Google que está disponible principalmente en dispositivos móviles y altavoces inteligentes. El Asistente de Google puede entablar conversaciones bidireccionales. Google Now vio la luz en mayo de 2016 como parte de su altavoz inteligente Google Home. Con posterioridad comenzó a incluirse en dispositivos Android en febrero de 2017.

Los usuarios interactúan principalmente con el Asistente de Google a través de la voz natural, aunque también se admite la entrada del teclado. El Asistente puede buscar en Internet, programar eventos y alarmas, ajustar la configuración del hardware en el dispositivo del usuario y mostrar información de la cuenta de Google del usuario. Google también ha anunciado que el Asistente podrá identificar objetos y recopilar información visual a través de la cámara del dispositivo, y admitir la compra de productos y el envío de dinero.

### 1.9.4 Alexa

Amazon Alexa es una tecnología de asistente virtual que emplea los servicios en la nube de Amazon para realizar el reconocimiento de voz y el procesamiento del lenguaje natural. En el año 2013, Amazon adquirió el software de reconocimiento de voz de Ivona que ha servido como base para el desarrollo de la tecnología de reconocimiento de voz de Alexa. Se ha incluido por primera vez en los altavoces inteligentes Amazon Echo y Echo Dot.

Es capaz de soportar conversaciones con seguimiento de contexto, reproducir música, leer libros y ofrecer información procedente de búsquedas en internet. También puede controlar dispositivos inteligentes y puede servir como base de un sistema domótico de automatización en el hogar. Mediante plugins puede extenderse su funcionalidad para realizar todo tipo de tareas.

## 1.10 Referencias